\documentclass[a4paper]{jpconf}

\usepackage{amsmath, amsfonts,
   amssymb, amsbsy, 
   euscript}

\begin{document}
\title{A convenient criterion under which $\mathbb{Z}_2$-\/graded operators are Hamiltonian}

\author{V\'eronique Hussin${}^1$ and Arthemy V Kiselev${}^{2,3}$}

\address{${}^1$ D\'epartement de Math\'ematiques et de Statistique,
Universit\'e de Montr\'eal,
C.P.~6128, succ.\ Centre\/-\/ville, Montr\'eal, Qu\'ebec H3C~3J7,
Canada.}

\address{${}^2$ J.~Bernoulli Institute for Mathematics and Computer Science, University of Groningen, P.O.Box~407, 9700~AK Gtoningen, The Netherlands.}
\address{${}^3$ IH\'ES, Le Bois\/-\/Marie~35, Route de Chartres,
Bures\/-\/sur\/-\/Yvette, 91440 France.}

\ead{hussin@dms.umontreal.ca, arthemy@ihes.fr} 

\begin{abstract}
We formulate a simple and convenient criterion under which skew\/-\/adjoint
$\mathbb{Z}_2$-\/graded total differential operators are Hamiltonian, provided that their images are closed under commutation in the Lie algebras of evolutionary vector fields on the infinite jet spaces for vector bundles over smooth manifolds.
\end{abstract}

\noindent%
In this short note we consider Hamiltonian differential operators that induce Poisson brackets on the spaces of Hamiltonian functionals on the infinite jet spaces for $\mathbb{Z}_2$-\/graded vector bundles over smooth (super-)\/manifolds. In other words, we study the structures that are related to the bundles in which the fibres are split, as vector spaces, in the even and the odd components so that, in particular, the even components of the local sections commute with everything whereas the odd components of the sections anti\/-\/commute between themselves. In addition, the base of the bundles can be a supermanifold itself, whence the super\/-\/derivatives emerge; we always assume that the operators at hand are polynomial in the (super-)\/derivatives.

We extend a very simple criterion (\cite{d3Bous} and
\cite[p.~130]{Opava}), under which linear 
differential operators 
are Hamiltonian, to the $\mathbb{Z}_2$-\/graded setup.
This will be helpful, in particular, in the study of supersymmetric integrable systems. 
Obviously, the same criterion allows us to check the compatibility~\cite{Magri}
of two given $\mathbb{Z}_2$-\/graded Hamiltonian operators $A_1,A_2$
by verifying that the linear combinations $A_1+\lambda A_2$ remain Hamiltonian at all~$\lambda$. 
The tool which we elaborate is 
very practical and efficient: indeed, its ``hardest'' component amounts to the calculation of the commutator of two 
evolutionary vector fields (c.f.~\cite{Topical}).
It is important that the procedure is purely algorithmic and is applicable immediately without any further adaptations (handled, e.g., by the
software~\cite{SsTools}).
We recall that other methods for checking whether a given operator is Hamiltonian are available from the literature: e.g.\ one can re\/-\/derive the algorithmic verification procedure from~\cite{JKKerstenVerbN=1} in the $\mathbb{Z}_2$-\/graded setup;
that concept is based on the use of variational poly
vectors which are already endowed with their own grading.
We finally recall that the book~\cite{Olver} contains another step\/-\/by\/-\/step 
verification procedure but (especially
in the $\mathbb{Z}_2$-\/graded case) in practice it is much more involved.

This note is structured as follows.
We first extend the 
criterion of~\cite{d3Bous} to bosonic super\/-\/fields and super\/-\/operators, see~\eqref{EqDogma}.
Theorem~1 
in section~\ref{SecZ2}
is our main result that covers the general setup of
$\mathbb{Z}_2$-\/graded fields. Its proof, which is given here in full
detail, is considerably simplified with respect to the one
in~\cite{d3Bous}.

All notions and constructions from the geometry of differential equations are standard (\cite{Olver} and~\cite{Opava,Dubr}).
We follow the notation of~\cite{d3Bous} which agrees with that of~\cite{Opava}
but initially covered only the non\/-\/graded case.
In the sequel, everything is real and $C^\infty$-\/smooth. 

\section{Bosonic super\/-\/fields and Hamiltonian super\/-\/operators}%
\label{SecFrob}\noindent%
Let $B^n\ni x=(x^1,\ldots,x^n)$~be an $n$-\/dimensional orientable manifold
and let $\pi\colon E^{m+n}\xrightarrow[F^m]{}B^n$ be a vector
bundle over~$B^n$ with
$m$-\/dimensional fibres~$F^m\ni u=(u^1,\ldots,u^m)$.
By $J^\infty(\pi)$ we denote the infinite jet space over~
$\pi$. 
We denote by~$u_\sigma$,
$|\sigma|\geq0$, its fibre coordinates.
We also denote by~$\mathfrak{g}$ the Lie algebra of evolutionary vector fields~$\partial_\varphi$ on $J^\infty(\pi)$ and by $\Omega$ the linear space of variational covectors, which contains the variational derivatives $\delta\mathcal{H}/\delta u$ of the
Hamiltonian functionals~$\mathcal{H}\in
\bar{H}
$ and which is dual to~$\mathfrak{g}$ with respect to the coupling
$\langle\,,\,\rangle$ that takes values in~$\bar{H}$.

Let $A\colon\Omega\to\mathfrak{g}$ be a total differential operator the image of which
is closed with respect to the commutation in~$\mathfrak{g}$,
\begin{equation}\label{EqDefFrob}  
[\text{im}\,A, \text{im}\,A]\subseteq\text{im}\,A.
\end{equation}
This is indeed so for $\mathbb{Z}_2$\/-\/graded Hamiltonian operators;
the criterion in Theorem~1
, see below, makes it clear
that condition~\eqref{EqDefFrob} is not superfluous for their definition.
Further examples of \emph{non\/-\/Hamiltonian} differential
operators, 
the images of which in the Lie algebras of evolutionary vector fields are subject to
the collective commutation closure but the domains of which are different from~$\Omega$, are studied in~\cite{d3Bous,SymToda}
and~\cite{Leiden}.


The operator~$A$ transfers the Lie algebra structure 
$[\,,\,]\bigr|_{\mathrm{im}\,A}$ to the skew\/-\/symmetric {
bracket} $[\,,\,]_A$ in the quotient $\text{dom} A/\ker A$,   
\begin{equation}\label{EqOplusBBoth}
\bigl[A(\boldsymbol{p}),A(\boldsymbol{q})\bigr]=A\bigl([\boldsymbol{p},\boldsymbol{q}]_A),\qquad \boldsymbol{p},\boldsymbol{q}\in\Omega.
\end{equation}
By the Leibniz rule, two sets of summands appear in the bracket
$\bigl[\partial_{A(\boldsymbol{p})},\partial_{A(\boldsymbol{q})}\bigr]=\partial_{[A(\boldsymbol{p}),A(\boldsymbol{q})]}$
of evolutionary vector fields $\partial_{A(\boldsymbol{p})}$ and
$\partial_{A(\boldsymbol{q})}$: 
\begin{equation}\label{LeibnizForOp}
\bigl[A(\boldsymbol{p}),A(\boldsymbol{q})\bigr]=A\bigl(\partial_{A(\boldsymbol{p})}(\boldsymbol{q})-\partial_{A(\boldsymbol{q})}(\boldsymbol{p})\bigr)
  +\bigl(\partial_{A(\boldsymbol{p})}(A)(\boldsymbol{q})-\partial_{A(\boldsymbol{q})}(A)(\boldsymbol{p})\bigr).
\end{equation}
In the first term we have used the permutability of
evolutionary derivations, which are of the form
$\partial_\varphi=\varphi\,\tfrac{\partial}{\partial u}
+\tfrac{\mathrm{d}}{\mathrm{d} x}(\varphi)\,\tfrac{\partial}{\partial u_x}+\cdots$,
and total derivatives.
The second term hits the image of~$A$ by construction.
Consequently, the 
Lie algebra structure $[\,,\,]_A$ on the domain of~$A$
equals
\begin{equation}\label{EqOplusBKoszul} 
[\boldsymbol{p},\boldsymbol{q}]_A=\partial_{A(\boldsymbol{p})}(\boldsymbol{q})-\partial_{A(\boldsymbol{q})}(\boldsymbol{p})+ \{\!\{\boldsymbol{p},\boldsymbol{q}\}\!\}_A.
\end{equation}

\medskip
\noindent\textbf{Example 1.}\quad
The second Hamiltonian operator
for the Korteweg\/--\/de Vries equation is
$A=-\tfrac{1}{2}\,\tfrac{\mathrm{d}^3}{\mathrm{d} x^3}+
u\,\tfrac{\mathrm{d}}{\mathrm{d} x}+\tfrac{\mathrm{d}}{\mathrm{d} x}\circ u$,
where $\tfrac{\mathrm{d}}{\mathrm{d} x}=\tfrac{\partial}{\partial x}+u_x\,\tfrac{\partial}{\partial u}+\cdots$.
The image of~$A$ is closed under commutation, and the Lie algebra
structure $[\,,\,]_A$ on its domain is related by the 
homomorphisms $\delta/\delta u$ and~$A$ to the Lie algebra $\bigl(\bar{H},
\{\,,\,\}_A\bigr)$ of Hamiltonians, endowed with the Poisson bracket,
and to the Lie algebra $\bigl(\mathfrak{g},[\,,\,]\bigr)$ of evolutionary vector
fields, respectively (see~\cite{Dorfman}).
It can easily be checked~\cite{SokolovUMN} that,
for the above operator, the bracket $[\,,\,]_A$ on the domain~$\Omega$
of~$A$ 
is
$
[p,q]_A=\partial_{A(p)}(q)-\partial_{A(q)}(p)+\tfrac{\mathrm{d}}{\mathrm{d} x}(p)\cdot q
   -p\cdot\tfrac{\mathrm{d}}{\mathrm{d} x}(q)
$, here $p,q\in\Omega$. 

\medskip
The bracket $\{\!\{\,,\,\}\!\}_{A}$ for Hamiltonian operators~$A$ can be obtained
explicitly from the Jacobi identity
$\lshad \boldsymbol{A},\boldsymbol{A}\rshad=0$ for the Lie algebra
$\bigl(\bar{H} 
,\{\,,\,\}_A\bigr)$ of the Hamiltonian functionals
endowed by~$A$ with the Poisson bracket $\{\,,\,\}_A$;
here $\boldsymbol{A}$ is the representation of~$A$ by the variational Poisson
bi\/-\/vector and $\lshad\,,\,\rshad$ is the variational Schouten bracket,
see~\cite{d3Bous,Opava,Topical}. 
We now write the result of such a derivation in local coordinates
but in a properly ordered way which is slightly different from
Eq.~(5) in~\cite{d3Bous}: For
a 
Hamiltonian ope\-ra\-tor
$A=\bigl\|A^{ij}_{\boldsymbol{\tau}}\cdot\tfrac{\mathrm{d}^{|\boldsymbol{\tau}|}}{\mathrm{d}
x^{\boldsymbol{\tau}}}\bigr\|$,
the $k$-th ($1\leq k \leq m$)
component of~$\{\!\{\,,\,\}\!\}_{A}$ equals
\begin{equation}\label{EqDogma}
\{\!\{\boldsymbol{p},\boldsymbol{q}\}\!\}_{A}^k=\sum_{|\boldsymbol{\sigma}|,|\boldsymbol{\tau}|\geq 0}
\sum_{i,j=1}^m 
\Bigl(\frac{\mathrm{d}^{|\sigma|}}{\mathrm{d} x^\sigma}\Bigr)^{\dagger} \Bigl[
 q_i\cdot
   \frac{\partial A_{\boldsymbol{\tau}}^{ij}}{\partial u^k_{\boldsymbol{\sigma}}}\cdot
 \frac{\mathrm{d}^{|\boldsymbol{\tau}|}}{\mathrm{d} x^{\boldsymbol{\tau}}} (p_j) \Bigr],
\end{equation}
where $\dagger$ denotes the adjoint. 
The benefit of this notation is that formula~\eqref{EqDogma} covers
the super\/-\/setup of bosonic super\/-\/fields and parity\/-\/preserving Hamiltonian operators that endow the spaces of
bosonic functionals with Poisson brackets. 
Here the multi\/-\/indices $\boldsymbol{\sigma}$ and~$\boldsymbol{\tau}$ can run through the super\/-\/derivations as well, and the partial derivatives $\partial/\partial u^k_{\boldsymbol{\sigma}}$ in~\eqref{EqDogma} act according to the graded Leibniz rule.

\medskip
\noindent\textbf{Example 2.}\quad 
Let $\boldsymbol{u}=u_0(x,t)\cdot\mathbf{1}+\theta_1\cdot u_1(x,t)+\theta_2\cdot u_2(x,t)+
\theta_1\theta_2\cdot u_{12}(x,t)$ be a scalar bosonic super\/-\/field, that is,
a mapping of $\mathbb{R}^2\ni(x,t)$ to the four\/-\/dimensional Grassmann algebra generated over~$\mathbb{R}$ by $\theta_1$ and~$\theta_2$ satisfying $\theta_i\theta_j=-\theta_j\theta_i$. By definition, put $\mathcal{D}_i=\partial/\partial\theta_i+\theta_i\cdot\mathrm{d}/\mathrm{d} x$, here $1\leq
 i,j\leq 2$ and it is readily seen that~$\mathcal{D}_i\mathcal{D}_j+\mathcal{D}_j\mathcal{D}_i=2\delta_{ij}\cdot\mathrm{d}/\mathrm{d} x$.

Consider the super\/-\/operator~$\boldsymbol{A}_2$
that comes from the $\mathsf{N}{=}2$ classical super\/-\/conformal
algebra~\cite{ChaichianKulish87} and yields the second Hamiltonian structure
for the triplet of integrable $\mathsf{N}{=}2$ supersymmetric Korteweg\/--\/de Vries
equations (\cite{MathieuNew}, see also~\cite{DefA4})
\begin{equation}\label{N=2SecondHam}
\boldsymbol{A}_2=\mathcal{D}_1\mathcal{D}_2\tfrac{\mathrm{d}}{\mathrm{d} x}+2\boldsymbol{u}\tfrac{\mathrm{d}}{\mathrm{d} x}
   -\mathcal{D}_1(\boldsymbol{u})\mathcal{D}_1-\mathcal{D}_2(\boldsymbol{u})\mathcal{D}_2+2\boldsymbol{u}_x.
\end{equation}
Let the bosonic super\/-\/sections $\boldsymbol{p},\boldsymbol{q}\in\Omega$
be two arguments of~$\boldsymbol{A}_2$.
Then formula~\eqref{EqDogma} yields their skew\/-\/symmetric bracket
\begin{equation}\label{N2Bracket}
\{\!\{\boldsymbol{p},\boldsymbol{q}\}\!\}_{\boldsymbol{A}_2}=
2\bigl(\tfrac{\mathrm{d}}{\mathrm{d} x}\boldsymbol{p}\cdot\boldsymbol{q}-\boldsymbol{p}\cdot\tfrac{\mathrm{d}}{\mathrm{d} x}\boldsymbol{q}\bigr)
 -\mathcal{D}_1(\boldsymbol{p})\cdot\mathcal{D}_1(\boldsymbol{q})-\mathcal{D}_2(\boldsymbol{p})\cdot\mathcal{D}_2(\boldsymbol{q}),
\end{equation}
and the validity of~\eqref{EqOplusBKoszul}
confirms 
that the super\/-\/operator~$\boldsymbol{A}_2$ is indeed Hamiltonian.

\section{$\mathbb{Z}_2$\/-\/graded fields and the Hamiltonianity criterion}%
\label{SecZ2}\noindent%
The purely bosonic setup of~\cite{d3Bous,Opava} 
and the $\mathsf{N}{=}2$ supersymmetry invariance in Example~2 
are particular cases in the general $\mathbb{Z}_2$-\/graded framework of $(m_0 
\mid m_1)$-\/dimensional fibre 
bundles~$\pi$ and parity\/-\/preserving
Hamiltonian operators $A\colon\Omega\to\mathfrak{g}$
for 
bosonic Hamiltonian functionals. 

Let $\langle\,,\,\rangle$ denote the standard coupling
$\Omega\times\mathfrak{g}\to\bar{H}$ 
and define $\langle\,\mid\,\rangle$ by setting
$\langle\boldsymbol{b}\mid\boldsymbol{d}\rangle\mathrel{{:}{=}}\langle\boldsymbol{d},\boldsymbol{b}\rangle$.
Namely, if $\boldsymbol{b}=(\boldsymbol{b}^0,\boldsymbol{b}^1)$ and $\boldsymbol{d}=(\boldsymbol{d}^0,\boldsymbol{d}^1)$~are decomposed
to even and odd\/-\/graded components, then $\langle\boldsymbol{b},\boldsymbol{d}\rangle=\boldsymbol{b}^0
\cdot\boldsymbol{d}^0+\boldsymbol{b}^1\cdot\boldsymbol{d}^1$ and
$\langle\boldsymbol{b}\mid\boldsymbol{d}\rangle=\boldsymbol{b}^0\cdot\boldsymbol{d}^0-\boldsymbol{b}^1\cdot\boldsymbol{d}^1$. 
The definition of adjoint graded operators implies $\langle\boldsymbol{b},A(\boldsymbol{d})\rangle=
\langle\boldsymbol{d},A^\dagger(\boldsymbol{b})\rangle=\langle A^\dagger(\boldsymbol{b})\mid\boldsymbol{d}\rangle$.

\medskip
\noindent\textbf{Theorem 1.}\quad
\textit{A $\mathbb{Z}_2$-\/graded parity\/-\/preserving skew\/-\/adjoint total differential operator~$A\colon\Omega\to\mathfrak{g}$
is Hamiltonian if and only if its image is closed under commutation and, for all $\boldsymbol{p},\boldsymbol{q},\boldsymbol{r}\in\Omega
$, the bracket~$\{\!\{\,,\,\}\!\}_{A}$ in~\eqref{EqOplusBKoszul} satisfies the equality
\begin{equation}\label{NowByHands}
\bigl\langle A\bigl(\{\!\{\boldsymbol{p},\boldsymbol{q}\}\!\}_{A}\bigr) \mid \boldsymbol{r}\bigr\rangle
   = {:}\,\bigl\langle\boldsymbol{p},\partial_{\underleftarrow{A}(\underrightarrow{\boldsymbol{r}})}(A)(\boldsymbol{q})\bigr\rangle\,{:},
\end{equation}
where the normal order ${:}\ {:}$ suggests that all derivations are thrown off~$A(\boldsymbol{r})$ by the graded Green formula
and the arrows indicate that first $A(\boldsymbol{r})$~is moved to the right of
$\boldsymbol{q}$, and then the operator~$A$ is pushed to the left of 
$\boldsymbol{p}$ by Green's formula again
\textup{(}this is explained in the proof below\textup{)}. 
The arising argument of the skew\/-\/adjoint operator~$A$ is the bracket~$\{\!\{\boldsymbol{p},\boldsymbol{q}\}\!\}_{A}$.}

\medskip
\noindent\textit{Proof.}\quad
Let us expand each of the three terms of the Jacobi identity,
\[
\sum_{{\circlearrowright}} \partial_{A(\boldsymbol{p})}(\langle\boldsymbol{q},A(\boldsymbol{r})\rangle)=0,
\]
by using the Leibniz rule. We obtain
\begin{equation}\label{Jacobi3sumsNew}
\sum_{\circlearrowright}\Bigl[
\langle \partial_{A(\boldsymbol{p})}(\boldsymbol{q}),A(\boldsymbol{r})\rangle
 +\langle \boldsymbol{q},\partial_{A(\boldsymbol{p})}(A)(\boldsymbol{r})\rangle
 +\langle \boldsymbol{q}, A(\partial_{A(\boldsymbol{p})}(\boldsymbol{r}))\rangle \Bigr]=0.
\end{equation}
Consider the third term in~\eqref{Jacobi3sumsNew} and, by the substitution
principle~\cite{Olver}, suppose that $\boldsymbol{r}$~is the variational derivative of a Hamiltonian functional, whence 
the linearization~$\ell_{\boldsymbol{r}}$ is self\/-\/adjoint in the graded sense.
Consequently,
\begin{multline*}
\langle \boldsymbol{q}, A(\partial_{A(\boldsymbol{p})}(\boldsymbol{r}))\rangle
 = -\langle A(\boldsymbol{q}) \mid \partial_{A(\boldsymbol{p})}(\boldsymbol{r})\rangle
 = -\langle A(\boldsymbol{q}) \mid \ell_{\boldsymbol{r}}(A(\boldsymbol{p}))\rangle
 = -\langle A(\boldsymbol{p}) \mid \ell_{\boldsymbol{r}}^\dagger(A(\boldsymbol{q}))\rangle \\
 = -\langle A(\boldsymbol{p}) \mid \ell_{\boldsymbol{r}}(A(\boldsymbol{q}))\rangle
 = -\langle \ell_{\boldsymbol{r}} (A(\boldsymbol{q})), A(\boldsymbol{p})\rangle
 = -\langle \partial_{A(\boldsymbol{q})}(\boldsymbol{r}), A(\boldsymbol{p})\rangle.
\end{multline*}
Substituting this back in~\eqref{Jacobi3sumsNew} and
taking the sum over the cyclic permutations, we cancel 
$3\times 2$ terms, except for   
\begin{equation}\label{ExpandOneSum}
\langle\boldsymbol{q},\partial_{A(\boldsymbol{p})}(A)(\boldsymbol{r})\rangle
 + \langle\boldsymbol{r},\partial_{A(\boldsymbol{q})}(A)(\boldsymbol{p})\rangle
 + \langle\boldsymbol{p},\partial_{A(\boldsymbol{r})}(A)(\boldsymbol{q})\rangle=0.
\end{equation}
Now we consider separately the first and second summands in~\eqref{ExpandOneSum},
paying due attention to the order of graded objects and the directions the derivations act in. First, applying the even vector field $\partial_{A(\boldsymbol{p})}$ to the equality $\langle\boldsymbol{q},A(\boldsymbol{r})\rangle 
=\langle A^\dagger(\boldsymbol{q}) \mid \boldsymbol{r}\rangle$ and using $A^\dagger=-A$, we
conclude that
\[
\langle\boldsymbol{q},\partial_{A(\boldsymbol{p})}(A)(\boldsymbol{r})\rangle
 = -\langle\partial_{A(\boldsymbol{p})}(A)(\boldsymbol{q}) \mid \boldsymbol{r}\rangle.
\]
Likewise, the second summand in~\eqref{ExpandOneSum} gives
\[
\langle\boldsymbol{r},\partial_{A(\boldsymbol{q})}(A)(\boldsymbol{p})\rangle=\langle\partial_{A(\boldsymbol{q})}(A)(\boldsymbol{p})\mid\boldsymbol{r}\rangle.
\]
Hence from~\eqref{ExpandOneSum} we obtain
\[ 
\langle\partial_{A(\boldsymbol{p})}(A)(\boldsymbol{q})\mid\boldsymbol{r}\rangle
 - \langle\partial_{A(\boldsymbol{q})}(A)(\boldsymbol{p})\mid\boldsymbol{r}\rangle
=\langle\boldsymbol{p},\partial_{A(\boldsymbol{r})}(A)(\boldsymbol{q})\rangle. 
\]
Integrating the right\/-\/hand side by parts,
we move the skew\/-adjoint operator~$A$ off~$\boldsymbol{r}$
and obtain the bracket $\{\!\{\boldsymbol{p},\boldsymbol{q}\}\!\}_{A}$ as its argument. 

We have shown that if the bracket induced on the domain of
a given graded skew\/-\/adjoint operator~$A$ with involutive image,
see~\eqref{EqOplusBKoszul},
coincides with the bracket $\{\!\{\,,\,\}\!\}_{A}$ emerging from~\eqref{NowByHands},
then $A$~is indeed Hamiltonian, and \textit{vice versa}.
This concludes the proof.

\medskip
\noindent\textbf{Example 3.}\quad
Writing the super\/-\/operator~\eqref{N=2SecondHam} in components 
(see~\ref{AppPass} below), 
now with $p_i=\delta\mathcal{H}/\delta u_i$, whence $p_0$ and $p_{12}$ are even and $p_1$, $p_2$ are odd, we obtain the $(4\times 4)$-\/matrix
operator~\cite{ChaichianKulish87}
\begin{equation}\label{SecondHam4x4}
A_2=\begin{pmatrix}   
-\tfrac{\mathrm{d}}{\mathrm{d} x} & -u_2 & u_1 & 2u_0\tfrac{\mathrm{d}}{\mathrm{d} x}+2u_{0;x} \\
-u_2 & \bigl(\tfrac{\mathrm{d}}{\mathrm{d} x}\bigr)^2+u_{12} & -2u_0\tfrac{\mathrm{d}}{\mathrm{d} x}-u_{0;x} & 3u_1\tfrac{\mathrm{d}}{\mathrm{d} x}+2u_{1;x} \\
u_1 & 2u_0\tfrac{\mathrm{d}}{\mathrm{d} x}+u_{0;x} & \bigl(\tfrac{\mathrm{d}}{\mathrm{d} x}\bigr)^2+u_{12}\vphantom{\Bigr)} & 3u_2\tfrac{\mathrm{d}}{\mathrm{d} x}+2u_{2;x} \\
2u_0\tfrac{\mathrm{d}}{\mathrm{d} x} & -3u_1\tfrac{\mathrm{d}}{\mathrm{d} x}-u_{1;x} & -3u_2\tfrac{\mathrm{d}}{\mathrm{d} x}-u_{2;x} & \bigl(\tfrac{\mathrm{d}}{\mathrm{d} x}\bigr)^3+4u_{12}\tfrac{\mathrm{d}}{\mathrm{d} x}+2u_{12;x}
\end{pmatrix}.
\end{equation}
The application of Theorem~1 
is particularly transparent 
since the coefficients of~\eqref{SecondHam4x4} are linear functions. The right\/-\/hand side of~\eqref{NowByHands} yields the four components of the skew\/-\/symmetric
bracket~$\{\!\{\boldsymbol{p},\boldsymbol{q}\}\!\}_{A_2}$,
\begin{align*}
\{\!\{\boldsymbol{p},\boldsymbol{q}\}\!\}_{A_2}^0&=2(p_{0;x}q_{12}-p_{12}q_{0;x})-(p_{1;x}q_2+p_2q_{1;x})
+(p_{2;x}q_1+p_1q_{2;x}),\\
\{\!\{\boldsymbol{p},\boldsymbol{q}\}\!\}_{A_2}^1&=2(p_{1;x}q_{12}-p_{12}q_{1;x})+(p_0q_2-p_2q_0)
+(p_{12;x}q_1-p_1q_{12;x}),\\
\{\!\{\boldsymbol{p},\boldsymbol{q}\}\!\}_{A_2}^2&=2(p_{2;x}q_{12}-p_{12}q_{2;x})+(p_1q_0-p_0q_1)
+(p_{12;x}q_2-p_2q_{12;x}),\\
\{\!\{\boldsymbol{p},\boldsymbol{q}\}\!\}_{A_2}^{12}&=2(p_{12;x}q_{12}-p_{12}q_{12;x})-p_1q_1-p_2q_2.
\end{align*}
This is the component expansion of~\eqref{N2Bracket};
see~\cite{DefA4} 
for further results on the geometry of the $\mathsf{N}{=}2$
supersymmetric $a{=}4$-\/Korteweg\/--\/de Vries equation. 

\section*{Conclusion}
Theorem~1 
provides an exact and exhaustive answer on the question whether a given skew\/-\/adjoint $\mathbb{Z}_2$-\/graded
differential operator with involutive image in~$\mathfrak{g}$ is Hamiltonian:
\begin{itemize}
\item take sections~$\boldsymbol{p},\boldsymbol{q}\in\Omega$ and calculate the commutator
$\bigl[A(\boldsymbol{p}),A(\boldsymbol{q})\bigr]$, omitting the standard terms
$\partial_{A(\boldsymbol{p})}(\boldsymbol{q})-\partial_{A(\boldsymbol{q})}(\boldsymbol{p})$;
\item calculate the $m$-\/tuple $A\bigl(\{\!\{\boldsymbol{p},\boldsymbol{q}\}\!\}_{A}\bigr)$ by
using formula~\eqref{NowByHands}.
\end{itemize}
If the two expressions coincide, the operator~$A$ is Hamiltonian.

\appendix
\section{Field\/--\/superfield correlation for variational derivatives}\label{AppPass}\noindent%
Given a Hamiltonian functional $\boldsymbol{\mathcal{H}}=\int\boldsymbol{h}[\boldsymbol{u}]\,\mathrm{d}\boldsymbol{\theta}\mathrm{d} x$, $\mathrm{d}\boldsymbol{\theta}=\mathrm{d}\theta_1\cdot\ldots\cdot\mathrm{d}\theta_N$ whose density $\boldsymbol{h}$ is a differential superfunction in $
(m_0\mid m_1)$ super\/-\/fields~$\boldsymbol{u}^\alpha$ of $2^N$~components each, what is the correlation between the components $p_I^\alpha$ of the variational derivatives
\begin{align}
\frac{\delta\boldsymbol{\mathcal{H}}}{\delta\boldsymbol{u}^\alpha}&=p^\alpha_\varnothing\cdot\mathbf{1}+\ldots+
\theta_1\cdots\theta_N\cdot p^\alpha_{(1,\ldots,N)}
=\sum_{|I|=0}^N\boldsymbol{\theta}_I\cdot p^\alpha_I\notag
\\
\intertext{with respect to the super\/-\/fields}
\boldsymbol{u}^\alpha&=u^\alpha_\varnothing\cdot\mathbf{1}+\ldots+\theta_1\cdots\theta_N\cdot u^\alpha_{(1,\ldots,N)}
=\sum_{|J|=0}^N\boldsymbol{\theta}_J\cdot u^\alpha_J\label{SuperField}
\end{align}
and, on the other hand, the variational derivatives
$\psi^\alpha_J={\delta\boldsymbol{\mathcal{H}}}\bigr/{\delta u^\alpha_J}$
of the functional~$\boldsymbol{\mathcal{H}}$ with respect to the $(m_0+m_1)\cdot 2^N$ components~$u^\alpha_J$ of the super\/-\/fields 
(here $1\leq\alpha\leq m_0+m_1$)\,? 
We note that the answer to this question (for which it suffices to consider only one super\/-\/field, hence we shall omit the superscripts~$\alpha$)
also encodes the 
Hamiltonian super\/-\/operators in their matrix component form (e.g., see~\eqref{SecondHam4x4}). 

\medskip
\noindent\textbf{Proposition.}\quad
Let $\boldsymbol{u}
$~be an $N\geq1$ super\/-\/field~\eqref{SuperField} and 
$\boldsymbol{\mathcal{H}}=\int \boldsymbol{h}[\boldsymbol{u}]\,\mathrm{d}\boldsymbol{\theta}\mathrm{d} x$ be a Ha\-mil\-to\-ni\-an super\/-\/functional. For all
multi\/-\/indexes~$J$ of length $|J|$ such that $0\leq|J|\leq N$, 
denote by~$I$ the multiindex of length $|I|=N-|J|$ such that 
their disjoint union is $I\sqcup J=\{1,\ldots,N\}$.
Then the sought correlation between $p_I$ and $\psi_J$~is
\begin{equation}\label{Rel}
\psi_J=(-1)^{(|\boldsymbol{\mathcal{H}}|-|\boldsymbol{u}|-|I|)\cdot|J|}\cdot(-1)^{\overline{I,J}}\cdot p_I,
\end{equation}
where $|\boldsymbol{\mathcal{H}}|$~is the parity of the Hamiltonian, $|\boldsymbol{u}|$~is the parity of the super\/-\/field, and the ordered concatenation of the multi\/-\/indexes $\overline{I,J}$~is a permutation of~$1,\ldots,N$.

\medskip
\noindent\textbf{Example 4.}\quad
Suppose $\mathsf{N}{=}2$ as in Examples~2 
and~3
. Then the correlation between the component expansion 
$\boldsymbol{p}=p_0\cdot\mathbf{1}+\theta_1 p_1+\theta_2 p_2+\theta_1\theta_2 p_{12}$ of the variational derivative~$\delta\boldsymbol{\mathcal{H}}/\delta\boldsymbol{u}$ and the variations~$\psi_I=\delta\boldsymbol{\mathcal{H}}/\delta u_I$ with respect to the components~$u_I$, $I\in\{0,1,2,12\}$, of the 
super\/-\/field~$\boldsymbol{u}=u_0\cdot\mathbf{1}+\theta_1u_1+\theta_2u_2+\theta_1\theta_2u_{12}$ is given by the formula\footnote{Although formula~\eqref{CorrN2} is valid for all~$\boldsymbol{\mathcal{H}}$, it is particularly transparent for~$\boldsymbol{h}=\tfrac{1}{2}\boldsymbol{u}^2$ such that the identity $\tfrac{1}{2}\int\boldsymbol{u}^2\,\mathrm{d}\boldsymbol{\theta}\mathrm{d} x=\int\bigl(u_0u_{12}-u_1u_2\bigr)\,\mathrm{d} x$ yields $\boldsymbol{p}=\boldsymbol{u}$ and $\vec{\psi}={}^{\mathrm{t}}\bigl(u_{12},u_2,-u_1,u_0\bigr)$,
the superscript~${}^{\mathrm{t}}$ denoting the transposition.}
\begin{equation}\tag{\ref{Rel}${}'$}\label{CorrN2}
\psi_0=p_{12},\qquad \psi_1=p_2,\qquad \psi_2=-p_1,\qquad
\text{and}\quad\psi_{12}=p_0.
\end{equation}
We thus recover the $(4\times4)$-\/matrix operator~\eqref{SecondHam4x4}
by writing in components both $\boldsymbol{u}$ and the argument~$\boldsymbol{p}$ of the Hamiltonian super\/-\/operator~\eqref{N=2SecondHam}, by inserting the correlation~\eqref{CorrN2} for the components of~$\boldsymbol{p}$, and then reordering the columns of the matrix operator~$A_2$ so that its argument~$\vec{\psi}$ acquires the standard form~${}^{\mathrm{t}}\bigl(\psi_0,\psi_1,\psi_2,\psi_{12}\bigr)$. To let the notation of Example~3 
match Theorem~1
, we re\/-\/denote by~$p_i$ the variational derivatives~$\delta\boldsymbol{\mathcal{H}}/\delta u_i$.

\medskip
\noindent\textit{Proof of Proposition.}\quad
Consider the Hamiltonian $\boldsymbol{\mathcal{H}}=\int\boldsymbol{h}\,\mathrm{d}\boldsymbol{\theta}\mathrm{d} x$. Varying the super\/-\/field~$\boldsymbol{u}$ by~$\delta\boldsymbol{u}$, we throw all the derivatives off~$\delta\boldsymbol{u}$ using multiple integration by parts in~$x$, which yields $\int\boldsymbol{p}\cdot\delta\boldsymbol{u}\,\mathrm{d}\boldsymbol{\theta}$. Next, let us insert the expansions $\boldsymbol{p}=\sum_I\theta_{i_1}\cdots\theta_{i_{|I|}}\cdot p_I$ and $\delta\boldsymbol{u}=\sum_J\theta_{j_1}\cdots\theta_{j_{|J|}}\cdot u_J$ in this super\/-\/integral.
By its definition, 
only the coefficient of~$\theta_1\cdots\theta_N$ in the product $\boldsymbol{p}\cdot\delta\boldsymbol{u}$ contributes to the integral's value,
hence only the complementary multi\/-\/indexes $I\sqcup J=\{1,\ldots,N\}$ count. Pushing~$p_I$ through $\boldsymbol{\theta}^J$, we accumulate the sign $(-1)^{|p_I|\cdot|J|}$, where $|p_I|=|\boldsymbol{\mathcal{H}}|-|\boldsymbol{u}|-|I|$. Finally,  reordering the product $\boldsymbol{\theta}_I\cdot\boldsymbol{\theta}_J$ to $\theta_1\cdots\theta_N$, we obtain the sign of the permutation~$\overline{I,J}$.

\ack 
The authors thank J.~W.~van de Leur 
for helpful discussions and constructive criticisms.
This research is partially supported by NSERC (for~V.\,H.)
and NWO grants~B61--609 and VENI~639.031.623 (for~A.\,K.).
A part of this research was done while A.\,K. was visiting at 
Max Planck Institute for Mathematics (Bonn);
the financial support and hospitality of this institution and the IH\'ES are
gratefully acknowledged.

\bigskip
\noindent\textit{J.~Phys.\ Conf. Ser.}: 
Mathematical and Physical Aspects of Symmetry. 
Proc.\ 28th Int.\ colloq.\ on group\/-\/theoretical methods in Physics (July 26--30, 2010; Newcastle\/-\/upon\/-\/Tyne, UK).

\noindent\textit{Submitted}: October~28, 2010; \textit{accepted}: January~6, 2011.
\end{document}